\documentclass[fleqn,twoside]{article}

\usepackage{espcrc2}
\usepackage{amsmath,euscript,psfrag,amssymb,graphicx}

\readRCS
$Id: espcrc2.tex,v 1.2 2004/02/24 11:22:11 spepping Exp $
\usepackage{graphicx}
\usepackage[figuresright]{rotating}

\newcommand{\be}{\begin{equation}}
\newcommand{\ee}{\end{equation}}
\newcommand{\bea}{\begin{eqnarray}}
\newcommand{\eea}{\end{eqnarray}}

\newcommand{\order}{{\cal O}}

\def\nb{\bar{n}}

\def\A{{\EuScript A}}

\def\X{{\EuScript X}}

\allowdisplaybreaks

\title{\begin{flushright}\normalsize\vspace{-1cm}
{\textsf SLAC-PUB-10847} \\ 
{\textsf hep-ph/0411073} \\
\end{flushright}
Symmetry relations for heavy-to-light meson form factors at large recoil 
\thanks{ Invited talk presented at the 11th International Conference in Quantum ChromoDynamics (QCD04), 
5-9 Jul 2004, Montpellier, France, 
}}

\author{Richard J. Hill  \\
        Stanford Linear Accelerator Center, 
        Stanford University, \\ 
        Stanford, CA 94309, U.S.A.%
       }
       
\runtitle{}
\runauthor{}

\begin{document}

\begin{abstract}
The description of large-recoil heavy-to-light meson 
form factors is reviewed in the framework of soft-collinear effective theory. 
At leading power in the heavy-quark expansion, three classes of approximate symmetry relations
arise.  The relations are compared to experimental data for $D\to K^*$ 
and $D_s\to \phi$ form factors, and to light-cone QCD sum rule predictions  
for $B\to \pi$ and $B\to \rho$ form factors.  Implications for the extraction of $|V_{ub}|$ 
from semileptonic $B\to \rho$ decays are discussed. 
\vspace{1pc}
\end{abstract}

\maketitle

\section{Introduction}

Form factors describing heavy meson decays into energetic light mesons are an essential ingredient 
for extracting CKM parameters from experimental $B$ decay measurements.  
The QCD description in this kinematic regime
is complicated by the existence of multiple energy scales, and by the competition of different scattering 
mechanisms.  The methods of effective field theory can be used to disentangle the different energy scales, 
and to provide a systematic heavy-quark expansion.  The soft-collinear effective theory (SCET) has
been developed to accomplish this 
task~\cite{Bauer:2000yr,Bauer:2001yt,Chay:2002vy,Beneke:2002ph,Hill:2002vw},\cite{Becher:2003qh,Becher:2004ms}.   

This talk reviews the description of form factors in SCET. 
The following Sec.~\ref{sec:ff} outlines the representation of weak current operators in the 
effective theory, and the resulting form factor expressions.  
Sec.~\ref{sec:symmetry} introduces three classes of symmetry relations which emerge from the effective theory at different 
levels of approximation.  
These predictions are compared with existing experimental data on $D$ meson decays, and with light-cone QCD sum rules (LCSRs)
for $B$ decays. The implications for extracting $|V_{ub}|$ from semileptonic $B\to \rho$ decays are briefly discussed.
Sec.~\ref{sec:renorm} comments on possible scaling violations of the form factors relative to the naive $q^2$ 
dependence derived from power counting in perturbation theory.    
A summary is presented in Sec.~\ref{sec:conclusion}. 

\section{Form factors in SCET}
\label{sec:ff}

The decay of a heavy $B$-meson into an energetic light meson necessarily involves
the interplay of soft partons in the heavy meson, with momentum $p_s$ of order $\Lambda_{\rm QCD}$,  
and ``collinear'' partons in the light meson, which carry a large energy $E$ in the direction of 
the outgoing meson but
have small virtuality, $p_c^2 \sim \Lambda_{\rm QCD}^2$.  
The QCD amplitudes are parameterized by 
matrix elements of local currents, e.g. 
\be
\begin{aligned}
  &\langle \pi(p^\prime)|\bar{q}\gamma^\mu b|\bar{B}(p)\rangle \\
  &\quad = F_+(q^2)(p+p^\prime)^\mu + F_-(q^2)(p-p^\prime)^\mu \,,
\end{aligned}
\ee
where $q=p-p^\prime$.  SCET describes
the large-energy/heavy-quark expansion of such 
quantities in the limit $E\sim m_b \gg \Lambda_{\rm QCD}$.   

It is convenient to work with light-cone coordinates, 
$(n\!\cdot\! k, \bar{n}\!\cdot\! k, k_\perp)$, where in the $B$ rest frame (with velocity 
$v^\mu=(1,0,0,0)$), 
and with the light meson emitted 
in the $+z$ direction, $n^\mu=(1,0,0,1)$ and $\bar{n}^\mu=(1,0,0,-1)$.  
Upon integrating out hard momentum modes, 
$k^2\gtrsim m_b^2$, QCD currents are matched onto two types of SCET 
operators relevant at leading order:~%
\footnote{
The notation and conventions are as in \cite{Hill:2004if,Becher:2004kk}. 
See also the references before (\ref{eq:factorization}). 
}
\begin{multline}
\label{Jexpansion}
  \bar q\,\Gamma\,b
  = C^A_i(E)\,J^A_i  \\
  + \int_0^1\! du\,C^B_j(E,u)\,J^B_j(u) + \dots \,,
\end{multline}
where 
\be
\begin{aligned}
  J^A_i &= \bar{\X}(0) \Gamma^{A}_{i}  h(0) \,, \\
  J^B_j(u) &= \int\!{ds\over 2\pi}\, e^{-iu(2E)s} \bar{\X}(s\bar{n})\, \A_{\perp\mu}(0)\, \Gamma^{B\mu}_j \, h(0) \,.  
\end{aligned}
\ee
Here $h$ is the heavy quark field in HQET~\cite{Neubert:1993mb}.       
$\X$ and $\A$ are fermion and gluon 
fields in SCET containing ``hard-collinear'' momenta, $p_{hc}^2\sim (p_s+p_c)^2 \sim m_b\Lambda_{\rm QCD}$. 
In $J^B(u)$, $\X$ and $\A$ carry fractions $u$ and $1-u$, respectively, of the large-component momentum $\nb\cdot p=2E$.

Upon taking matrix elements, the $A$-type SCET currents yield the soft-overlap form factor 
contributions (evaluated at renormalization scale $\mu$), 
\be
\begin{aligned}
\label{eq:projectors}
  &   {\langle M(p^\prime)| \bar{\X}\,\Gamma\,h |\bar{B}_v\rangle \over 2E } \\
  &\quad
   \equiv -\zeta_M(E,\mu)\,\,{\rm tr} \left\{ \overline{\cal M}_M(n) \Gamma {\cal M}(v) \right\}  \,, 
\end{aligned}
\ee
where $M=P$, $V_\perp$ or $V_\parallel$ 
denotes a light pseudoscalar or vector meson (with transverse or longitudinal polarization). 
${\cal M}_M(n)$ and 
${\cal M}(v)$ are spinor wavefunctions for $M$ and $B$ 
corresponding to the large-energy and heavy quark limits.  
The single function $\zeta_M(E)$ in (\ref{eq:projectors}) 
describes all soft-overlap contributions
to form factors involving the same final-state meson~\cite{Charles:1998dr}. 

The $B$-type SCET currents in (\ref{Jexpansion}) yield factorizable 
hard-scattering contributions to the form factors. 
After integrating out hard-collinear modes in a second matching step, 
these contributions may be expressed in terms of a
perturbatively calculable hard-scattering kernel convoluted with 
 light-cone wavefunctions, $\phi_B$ and $\phi_M$,  for the heavy and light mesons, respectively. 

At leading order in the $\Lambda_{\rm QCD}/m_b$ expansion, 
the form factors are therefore expressed as the 
sum of two terms~\cite{Beneke:2000wa,Bauer:2002aj,Beneke:2003pa,Lange:2003pk}:
\be
\label{eq:factorization}
F_i^{B\to M}(E) = C^A_i(E,\mu) \zeta_M(E,\mu)  + \Delta F^{B\to M}_i(E) \,. 
\ee
The hard-scattering term is given by 
\be
\begin{aligned}
\label{eq:DeltaF}
  &\Delta F^{B\to M}_i(E) = {m_B f_B f_M(\mu) \over 8 E K_F(\mu)} 
   \int_0^\infty {d\omega\over \omega} \int_0^1\!du\, \\ 
  &\times \phi_B(\omega,\mu) \, \phi_M(u,\mu) \\
  &\times \int_0^1 du^\prime {\cal J}_\Gamma\left(u,u^\prime,\ln{2E\omega\over \mu^2},\mu\right) C^B_i(E,u^\prime,\mu)  \,.
\end{aligned}
\ee
Here ${\cal J}_\Gamma$ is the Wilson coefficient for the second step of matching, with ${\cal J}_\Gamma={\cal J}_\parallel$
for $M=P,V_\parallel$ and ${\cal J}_\Gamma={\cal J}_\perp$ for $M=V_\perp$~\cite{Hill:2004if}.  
$f_B$ and $f_M$ are decay constants, and 
$K_F=1+\order(\alpha_s)$ relates the QCD and HQET heavy-meson decay constants~\cite{Neubert:1993mb}. 
The perturbative expansions of $C_i^B$ and of ${\cal J}_\Gamma$ in (\ref{eq:DeltaF}) involve logarithms of the ratios 
$\mu/E$ and $\mu^2/2E\omega$, with $\omega\sim \Lambda_{\rm QCD}$, so that   
large logarithms are unavoidable in fixed-order perturbation theory. 
These logarithms may be resummed using renormalization-group (RG) 
methods~\cite{Bauer:2000yr,Becher:2003kh,Lange:2003ff,Hill:2004if}.  

\section{Symmetry relations}
\label{sec:symmetry}

Eqs.~(\ref{eq:factorization}) and (\ref{eq:DeltaF}) can be used to relate any two $B\to M$ form factors.~%
\footnote{
Form factor conventions are as in \cite{Beneke:2000wa,Hill:2004if}. 
}
Given the 
meson LCDAs, these relations are perturbatively calculable up to $\Lambda_{\rm QCD}/m_b$ corrections.  However, since 
the $B$-meson LCDA is poorly constrained at present, it is useful to find relations which are independent of detailed
assumptions for $\phi_B(\omega)$.   Three types of relations arise, which for descriptive purposes will be called
``first-'', ``second-'' and ``third-class''.  

Two relations describing 
$B\to V_\perp$ decays hold to all orders in $\alpha_s$~\cite{Hill:2004if,Beneke:2000wa,Burdman:2000ku},
\be
\label{eq:firstclass}
\begin{aligned}
&\mbox{\rm First class relations:} \\
  A_1(q^2) &= \left( 1 - {\hat{q}^2}\right)\left(1+\hat{m}_V\right)^{-2} V(q^2) \,, \\
  T_2(q^2) &= \left( 1 - \hat{q}^2\right) T_1(q^2) \,.
\end{aligned}
\ee
Here $q\equiv p-p^\prime$ for the $B(p)\to M(p^\prime)$ transition, and hatted variables are 
in units of $m_B$: $\hat{m}_V=m_V/m_B$, $\hat{q}^2=q^2/m_B^2$, etc.  Kinematic factors linear in the light
mass are retained, and $q^2=m_B^2-2Em_B+\order(m_V^2)$ is used to express the form factors as a function of $q^2$.   

At tree level the coefficients $C^B_i$ in (\ref{eq:DeltaF}) are independent of momentum fraction $u^\prime$. 
When radiative corrections at the hard scale are neglected, the convolutions then yield a universal function $H_M(E)$~\cite{Hill:2004if},  
\be
\label{eq:DeltaFuniv}
  \Delta F_i^{B\to M}(E) \approx -\left(m_B\over 2E\right)^2 H_M(E) \, C_i^{B(\rm tree)}(E) \,.
\ee
Eq.~(\ref{eq:DeltaFuniv}) is exact to all orders in the perturbative expansion of 
the jet function ${\cal J}_\Gamma$, 
neglecting only hard-scale radiative corrections in $C_i^B$. 
In this approximation, the two hadronic functions,  $\zeta_M$ and $H_M$, may be eliminated to yield relations
between the three $B\to P$ and $B\to V_\parallel$ form factors,    
\be
\begin{aligned}
\label{eq:secondclass} 
  & \quad \qquad \mbox{\rm Second class relations:} \\
  &F_+(q^2) = F_0(q^2) + \left(1+\hat{m_P}\right)^{-1} \hat{q}^2 F_T(q^2) \,, \\
  &\left( 1+ \hat{m}_V \right)^{-1}\left[ V(q^2)-A_2(q^2) \right]  \\
  &\quad = 2\hat{m}_V A_0(q^2) + \hat{q}^2 \left[ T_1(q^2)-T_3(q^2) \right] \,.
\end{aligned}
\ee

Finally, neglecting the hard-scattering terms $\Delta F_i$ in (\ref{eq:factorization}) altogether yields the
``large-energy'' symmetry relations obeyed by the soft-overlap terms~\cite{Charles:1998dr}, 
\be
\begin{aligned}
\label{eq:thirdclass}
  &\mbox{\rm Third class relations:} \\
  F_+(q^2) &= \left(1-\hat{q}^2\right)^{-1} F_0(q^2) \\
  &= \left(1+\hat{m_P}\right)^{-1} F_T(q^2) \,,\\
  A_0(q^2) &= \left(1-\hat{q}^2 \right)(2\hat{m}_V)^{-1}\left(1+\hat{m}_V\right)^{-1} \\
  & \qquad \times \left[ V(q^2)-A_2(q^2) \right] \\
  &= (2m_V)^{-1}\left(1-\hat{q}^2\right)\left[ T_1(q^2)-T_3(q^2) \right] \,, \\
  T_1(q^2)  &= \left( 1+\hat{m}_V\right)^{-1} V(q^2)  \,. 
\end{aligned}
\ee

All relations in (\ref{eq:firstclass}), (\ref{eq:secondclass}) and (\ref{eq:thirdclass}) 
are expected to receive $\order(\Lambda_{\rm QCD}/E)$ corrections, of order $10-20\%$. 
Symmetry-breaking corrections in (\ref{eq:secondclass}) and (\ref{eq:thirdclass}) due to $C^A_i$ could be 
included trivially,
but this effect is $\lesssim 5\%$ in all cases~\cite{Beneke:2000wa,Becher:2004kk}.  
From current estimates of the hard-scattering terms,
$H/\zeta \sim 0.1-0.2$,~%
\footnote{
Estimates are based on LCSR form factor predictions (cf. Sec.~\ref{sec:LCSR}), or directly 
from sum rule analyses of the $B$-meson wavefunction~\cite{Grozin:1996pq,Braun:2003wx}.    
} 
radiative corrections to $C^B_j$ also have very little effect on symmetry relations. 
The second-class relations (\ref{eq:secondclass}) should then hold with similar accuracy to the first-class ones 
(\ref{eq:firstclass}), whereas the third-class
relations (\ref{eq:thirdclass}) ignore the hard-scattering terms entirely and so may receive larger corrections.   

\subsection{D decays}
\label{sec:Ddecay}

\begin{table}[t]
\begin{center}
\caption{Experimental values for form factor ratios $r_V$, $r_2$ and $r_3$ (see text), 
taken from \cite{Eidelman:2004wy}.}
\label{table:1}
\renewcommand{\tabcolsep}{1pc} % enlarge column spacing
\renewcommand{\arraystretch}{1.2} % enlarge line spacing
\begin{tabular}{lll}
\hline
      &  $D\to K^*$    & $D_s \to \phi$ \\
\hline
$r_V$ &  $1.62 \pm 0.08$ & $1.92 \pm 0.32$ \\
$r_2$ &  $0.83 \pm 0.05$ & $1.60 \pm 0.24$ \\
$r_3$ &  $0.04 \pm 0.33 \pm 0.29$ &  --- \\
\hline
\end{tabular}
\end{center}
\vspace{-8mm}
\end{table}

$D$ mesons, to the extent that the heavy quark expansion
in $\Lambda_{\rm QCD}/m_c$ is valid, can be analyzed in precisely the same way as $B$ mesons.  
Table~\ref{table:1} lists current experimental data for $r_V\equiv V(0)/A_1(0)$, $r_2\equiv A_2(0)/A_1(0)$ 
and $r_3\equiv \tilde{A}_3(0)/A_1(0)$.~%
\footnote{
$\tilde{A}_3 \equiv A_2/2 + \hat{m}_V(1+\hat{m}_V)(A_0-A_3)/\hat{q}^2$,
with $2\hat{m}_V A_3 = (1+\hat{m}_V) A_1 - (1-\hat{m}_V)A_2$.  This form factor contributes 
only when lepton masses are relevant. 
}
For instance, $r_V$
is determined by a first-class relation, (\ref{eq:firstclass}),   
which in terms of energy becomes, 
\be
\label{eq:symmE}
  2 \hat{E}\,V(E) = \left(1+\hat{m_V}\right)^2 A_1(E) \,. 
\ee
A similarity in measured values $r_V$ for $D\to K^*$ 
and $D_s\to \phi$ could suggest that this ratio is determined largely by
kinematic factors, as would be the case if a relation such as (\ref{eq:symmE}) were valid. 
In contrast, the ratio $r_2$ appears to exhibit a large $SU(3)$ symmetry-breaking.   
At $q^2=0$ for $D\to K^*$ decay, (\ref{eq:symmE}) yields 
\be
\label{eq:exrV}
  r_V^{D\to K^*}\! \approx\! (m_D+m_{K^*})^2/(m_D^2+ m_{K^*}^2)\! =\! 1.78 \,.
\ee

\begin{figure}[t]
\begin{center}
\psfrag{x}{\small $\hat{m}_V$}
\psfrag{y}{\small $r_V$}
\psfrag{f1}{\tiny $(1+\hat{m}_V)^2/(1+\hat{m}_V^2)$}
\psfrag{f2}{\tiny $(1+\hat{m}_V)/(1-\hat{m}_V)$}
\psfrag{f3}{\tiny $(1+\hat{m}_V)^2$}
\includegraphics*[width=18pc, height=10pc]{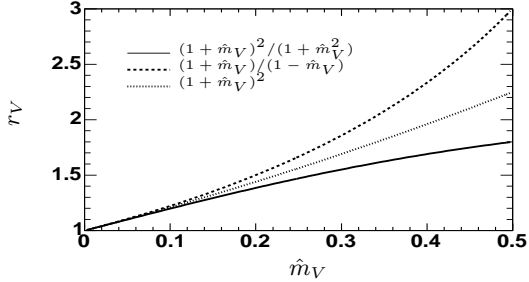}
\vspace{-16mm}
\caption{
Predictions for $r_V$ using the relations
$(1+\hat{m}_V)^{2} A_1/V = \{\, 2\hat{E}$ (solid), $2\hat{k}_V$ (dashed),  $\hat{E}+\hat{k}_V$ (dotted) \}.  
}
\vspace{-7mm}
\label{fig:r}
\end{center}
\end{figure}

The apparent agreement of (\ref{eq:exrV}) with the experimental value 
was counted as an early success of the large-energy
symmetry relations~\cite{Charles:1998dr}. 
However, before taking this agreement seriously, it is important to realize
that quadratic meson-mass effects shift the prediction considerably.  
For instance, directly from the definitions of form factors $V$ and $A_1$ in terms of 
QCD matrix elements, it is most natural to assume the symmetry relation, 
\be
\label{eq:symmkV}
  2 \hat{k}_V V(q^2) = (1+\hat{m}_V)^2 A_1(q^2) \,,
\ee
where $k_V\equiv\sqrt{E^2-m_V^2}=(\bar{n}\cdot p^\prime -n\cdot p^\prime)/2$ is the $3$-momentum of the light meson.  In this
case, 
\be 
  r_V^{D\to K^*} \approx (m_D+m_{K^*})/(m_D-m_{K^*}) = 2.81 \,.
\ee
Alternatively, it is convenient to express SCET quantities in terms of the  
large momentum component,  
$\nb\cdot p^\prime = E+k_V$.~%
\footnote{ In \cite{Hill:2004if,Becher:2004kk}, $\nb\cdot p^\prime$ is simply denoted by $2E$.  
}
The quantities $E$, $k_V$, and $(E+k_V)/2$ differ by terms 
of order $m_V^2/E^2$, and coincide in the large-energy limit.   
As shown in Fig.~\ref{fig:r}, the resulting difference 
in symmetry predictions for $r_V$ 
is very small for $B$ decays to light mesons, e.g. at $\hat{m}_V = m_{\rho}/m_B=0.15$, but is 
large for typical $D$ decays, e.g. at $\hat{m}_V=m_{K^*}/m_D=0.48$.  Unless such 
effects can be reliably accounted for, the application of large-energy
symmetry relations to $D$ decays appears problematic.  
This issue could be further investigated experimentally, 
for instance by measuring $r_V(q^2)\equiv V(q^2)/A_1(q^2)$
and comparing to the $q^2$ dependences predicted by (\ref{eq:symmE}) or (\ref{eq:symmkV}).  

SCET symmetry relations may also be used to relate $B$ and $D$ decay form factors involving the same final 
state meson.  For example, in decays to pseudoscalar mesons, at the level of second-class relations, 
\be
\label{eq:SCETgenP}
\begin{aligned}
  {F_+^{B\to P}\over F_+^{D\to P}} &= \sqrt{m_B\over m_D}{ \hat{\zeta}_P + \left( {4E\over m_B}-1\right) \hat{H}_P 
  \over \hat{\zeta}_P + \left( {4E\over m_D}-1\right) \hat{H}_P } \,.
\end{aligned}
\ee
Here $\hat{\zeta}_M$ and $\hat{H}_M$ are quantities independent of the heavy-quark mass, 
\be
  \zeta_M \equiv  \sqrt{m_B}\hat{\zeta}_M, \quad 
  \left(m_B\over 2E\right)^2 H_M \equiv \sqrt{m_B}\hat{H}_M \,.
\ee
Similarly, for decays to vector mesons,
\be
\label{eq:SCETgen}
\begin{aligned}
  {V^{B\to V}\over V^{D\to V}} &= \sqrt{m_D\over m_B}{m_B+m_V\over m_D+m_V} \,,\\
  {A_1^{B\to V}\over A_1^{D\to V}} &= \sqrt{m_B\over m_D}{m_D+m_V\over m_B+m_V} \,, \\
  { (V\!-\!A_2)^{B\to V}\over (V\!-\!A_2)^{D\to V}}\! &=\! \\
&\hspace{-20mm} \sqrt{m_D\over m_B}{m_B+m_V\over m_D+m_V}
  { \hat{\zeta}_{V_\parallel}\! +\! \left(\! {4E\over m_B}\!-1\!\right) \hat{H}_{V_\parallel} 
  \over \hat{\zeta}_{V_\parallel}\! +\! \left(\! {4E\over m_D}\!-1\!\right) \hat{H}_{V_\parallel} } \,.
\end{aligned}
\ee
Relations (\ref{eq:SCETgenP}) and (\ref{eq:SCETgen}) hold at the same value of the light meson energy $E$,
or equivalently 
at the same value of the recoil parameter $v\cdot v^\prime=E/m_M$, where $p^\prime = m_M v^\prime$. 
These are generalizations of corresponding relations in HQET, which counts $v\cdot v^\prime$ as order unity; 
when hard-scattering corrections $H_M$ are neglected,  the 
leading order predictions of HQET~\cite{Isgur:1990kf} are recovered.
In contrast, the SCET power counting allows $v\cdot v^\prime=E/m_M$ to be a large parameter. This is not 
very relevant for $D$ decays to vector mesons, e.g. $v\cdot v^\prime < 1.3$ in $D\to K^*$ decays, but 
is important near maximum recoil in $D\to \pi$ decays, where $v\cdot v^\prime \approx 6$.     

It is possible that the heavy-quark symmetries relating $B$ and $D$ decays (\ref{eq:SCETgen}) 
might still be valid, 
even if the large-energy symmetry predictions applied directly to the $D$ system, as 
in (\ref{eq:symmE}) and (\ref{eq:symmkV}), are not useful.      
Using LCSR predictions for $B$ and $B_s$ decays~\cite{Ball:1998kk}, and neglecting hard-scattering terms,  
the predictions from (\ref{eq:SCETgen}) 
for the ratios $r_V$, $r_2$ and $r_3$ for $D\to K^*$ are 1.71, 0.75 and 0.38, respectively, 
while the predictions for the ratios $r_V$ and $r_2$ for  $D_s \to \phi$ are 1.74 and 0.87.
These numbers are expected to receive $\sim 30\%$ corrections 
proportional to $1/m_c$, in addition to the $\gtrsim 15\%$ uncertainties from the $B$ decay form factors. 
The agreement with experimental values in Table~\ref{table:1} is reasonable with the possible exception
of $r_2^{D_s\to\phi}$.   
It would be interesting to test these relations more precisely when more data becomes available.

\subsection{Semileptonic branching fractions}

The differential rate for semileptonic $B$ decay to a vector meson (e.g. $V=\rho$) is,
neglecting lepton masses, 
\begin{multline}
  {d\Gamma(\bar{B}^0 \to V^+ l^- \bar{\nu} )\over d\hat{q}^2 d \cos\theta}  = 
  |V_{ub}|^2{G_F^2 m_B^5 \hat{k}_V \hat{q}^2\over 128\pi^3}  \\
  \times \bigg[ (1-\cos\theta)^2{|H_+|^2\over 2m_B^2}  + (1+\cos\theta)^2{|H_-|^2\over 2m_B^2} \\
  + \sin^2\theta {|H_0|^2\over m_B^2} \bigg] \,,
\end{multline}
where $\theta$ is the angle between the charged lepton in the virtual $W$ rest frame, and the 
direction of the $W$ in the $B$ rest frame.
The helicity amplitudes may be expressed in terms of
form factors, 
\be
\begin{aligned}
  {H_\pm\over m_B} &= { (1+\hat{m}_V)^2 A_1 \mp 2 \hat{k}_V V  \over 1+\hat{m}_V }  \,, \\
  {H_0\over m_B} &= { (1+\hat{m}_V)^2\left( \hat{E} - \hat{m}_V^2 \right) A_1 - 2\hat{k}_V^2 A_2 \over  
    \hat{m}_V(1+\hat{m}_V)\sqrt{\hat{q}^2} } \,, 
\end{aligned}
\ee
where as usual, hatted variables are in units of $m_B$.  
$H_+$ vanishes
at leading order in $\Lambda_{\rm QCD}/E$,  by the first-class symmetry relation (\ref{eq:symmkV}).~%
\footnote{
Note that (\ref{eq:symmkV}) is the ``exact'' form of the symmetry relation (cf. the discussion in
Sec.~\ref{sec:Ddecay}). In the same way, 
$H_0$ involves the ``exact'' version of the combination 
proportional to $V-A_2$.  
}   
In the large energy limit, the remaining helicity amplitudes are~%
\footnote{
Corrections to the large-energy limit can in principle be computed, but for simplicity they are not
included in the following discussion. 
} 
\be
  {H_-\over m_B} = 2\left(1-\hat{q}^2\right)\zeta_{V_\perp} \,,
  {H_0\over m_B} = {1\over \sqrt{\hat{q}^2}}\left(1-\hat{q}^2\right) \zeta_{V_\parallel} \,,
\ee
so that the differential rate satisfies 
\begin{multline}
\label{eq:diffpropto}
  {d\Gamma(B\to V l \nu)\over d\hat{q}^2} \propto
  (1-\hat{q}^2)^3 \int_{-1}^1 \! d\cos\theta \\ 
  \times \bigg[
  2(1+\cos\theta)^2\hat{q}^2 |\zeta_{V_\perp}|^2 + \sin^2\theta |\zeta_{V_\parallel}|^2 \bigg] \\
  = (1-\hat{q}^2)^3 \bigg[ 
  4\hat{q}^2 |\zeta_{V_\perp}|^2 + |\zeta_{V_\parallel}|^2 \bigg] \,.
\end{multline}
It is apparent from (\ref{eq:diffpropto}) that without angular discrimination, the 
contribution from $|\zeta_{V_\perp}|^2$ is suppressed relative to that
from $|\zeta_{V_\parallel}|^2$ at $\hat{q}^2 \lesssim 0.25$, or $q^2\lesssim 7\,{\rm GeV}^2$.   
Unfortunately, it is $\zeta_{V_\perp}$ that can be most cleanly constrained
by other measurements.
For example, the value of $\zeta_{V_\perp}$ at $q^2=0$ can be related to  
the $B\to\rho\gamma$ branching fraction.~%
\footnote{
If corrections to $SU(3)$ symmetry can be 
brought under control, the branching fractions for $B\to K^*\gamma$ and 
$B\to K^* l^+l^-$ could also be used to obtain both $\zeta_{V_\perp}$ and
$\zeta_{V_\parallel}$.
}
Reducing the uncertainty due to $\zeta_{V_\parallel}$ requires a restriction either to 
small $\theta$ or to larger values of $q^2$.  
The latter case, using $\hat{q}^2 \sim 0.6 -0.7$, has been proposed to extract 
$|V_{ub}|$ using constraints imposed by $D\to K^*$ and $D\to \rho$ decays~\cite{Ligeti:1999th}, although
a good understanding of power corrections 
is required for this approach to yield a precision measurement. 

Using the naive scaling prediction 
$\zeta_{V_\perp} \propto \zeta_{V_{\parallel}} \propto (1-\hat{q}^2)^{-2}$ in (\ref{eq:diffpropto})
gives some indication of the small $q^2$ behavior of $d\Gamma/d\hat{q}^2$:
\begin{multline}
{d\Gamma \over d\hat{q}^2} \propto
 1  + \left( 4{|\zeta_{V_\perp}|^2\over |\zeta_{V_\parallel}|^2} - 1\right)\hat{q}^2 + \dots \,,
\end{multline}
where $|\zeta_{V_\perp}|^2/|\zeta_{V_\parallel}|^2$ is evaluated at $q^2=0$.  
As discussed in the next section, 
LCSRs suggest that this ratio is close to unity, but also that
the scaling of $\zeta$ could receive significant corrections. 
Determining this residual $q^2$ dependence from first principles is an important problem.   For example, 
when combined with 
the extraction of $\zeta_{V_\perp}\!(q^2=0)$ from $B\to\rho\gamma$, it could provide a largely model-independent 
determination of $|V_{ub}/V_{td}V_{tb}|$~\cite{Bosch:2004nd}.  

\subsection{Light cone sum rules}
\label{sec:LCSR}

\begin{figure}[h]
\begin{center}
\psfrag{x}{\small $\hat{q}^2$}
\psfrag{y}{\small \hspace{-0.8cm} $(1-\hat{q}^2)^2 F(q^2)$}
\psfrag{f1}{\tiny $F_+$}
\psfrag{f2}{\tiny $F_0/(1-\hat{q}^2)$}
\psfrag{f3}{\tiny $F_T/(1+\hat{m}_\pi)$}
\includegraphics*[width=18pc, height=10pc]{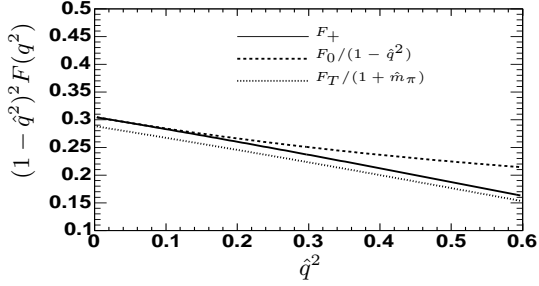}
\vspace{-16mm}
\caption{LCSR predictions for $B\to \pi$ form factors, taken from \cite{Ball:1998tj}}
\vspace{-8mm}
\label{fig:P}
\end{center}
\end{figure}
\begin{figure}[h]
\begin{center}
\psfrag{x}{\small $\hat{q}^2$}
\psfrag{y}{\small \hspace{-0.8cm} $(1-\hat{q}^2)^2 F(q^2)$}
\psfrag{f1}{\tiny $A_0$}
\psfrag{f2}{\tiny $(1-\hat{q}^2) (V-A_2)/(1+\hat{m}_\rho)/2\hat{m}_\rho$}
\psfrag{f3}{\tiny $(1-\hat{q}^2)(T_1-T_3)/2\hat{m}_\rho$}
\includegraphics*[width=18pc, height=10pc]{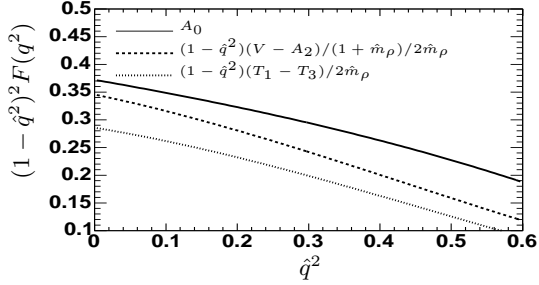}
\vspace{-16mm}
\caption{LCSR predictions for $B\to \rho_\parallel$ form factors, taken from \cite{Ball:1998kk}.}
\vspace{-7mm}
\label{fig:Vpar}
\end{center}
\end{figure}
\begin{figure}[h]
\begin{center}
\psfrag{x}{\small $\hat{q}^2$}
\psfrag{y}{\small \hspace{-0.8cm} $(1-\hat{q}^2)^2 F(q^2)$}
\psfrag{f1}{\tiny $(1+\hat{m}_\rho)A_1/(1-\hat{q}^2)$}
\psfrag{f2}{\tiny $V/(1+\hat{m}\rho)$}
\psfrag{f3}{\tiny $T_1$}
\psfrag{f4}{\tiny $T_2/(1-\hat{q}^2)$}
\includegraphics*[width=18pc, height=10pc]{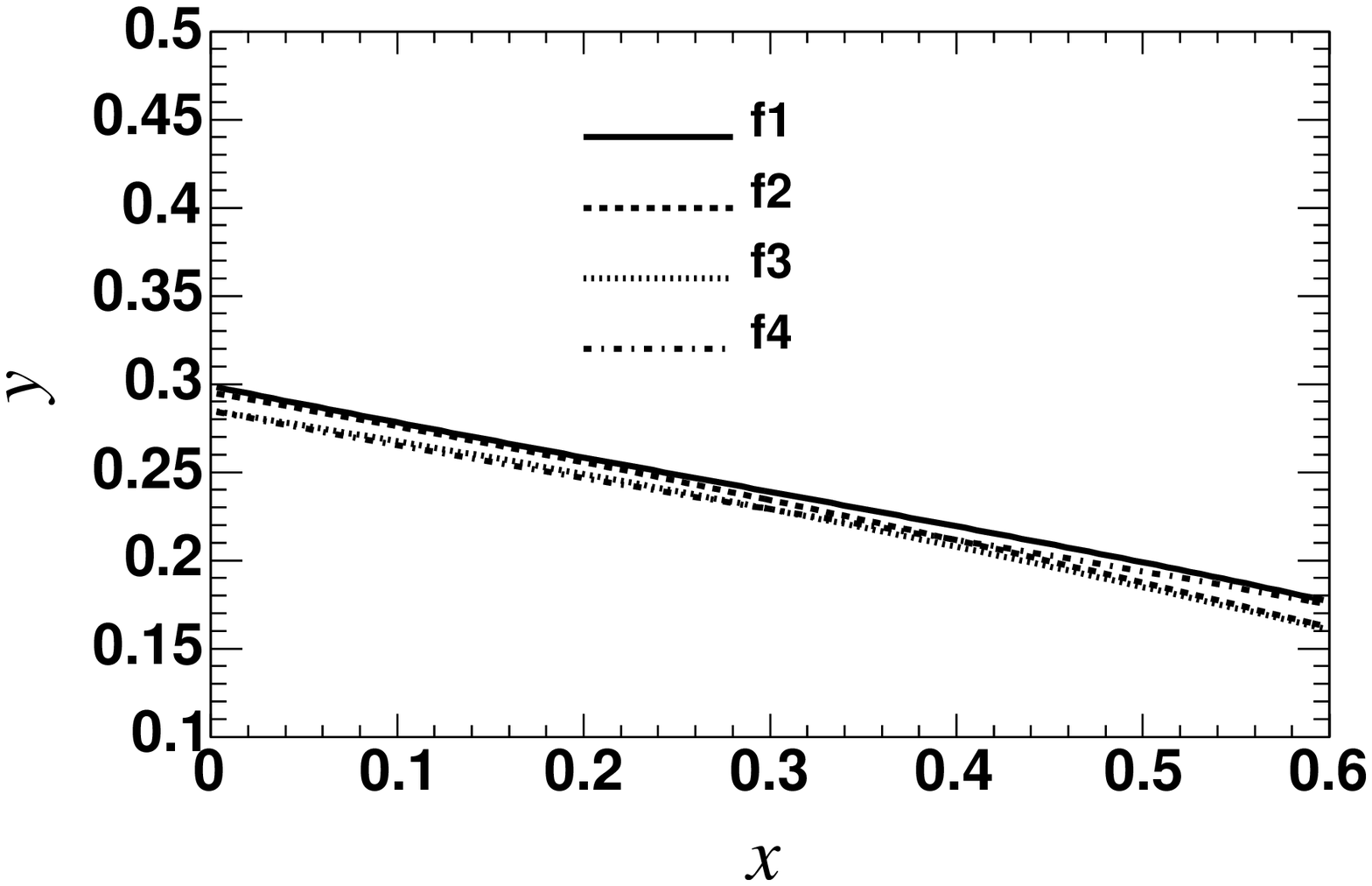}
\vspace{-16mm}
\caption{LCSR predictions for $B\to \rho_\perp$ form factors, taken from \cite{Ball:1998kk}.}
\vspace{-8mm}
\label{fig:Vperp}
\end{center}
\end{figure}
\begin{figure}[h]
\begin{center}
\psfrag{x}{\small $\hat{q}^2$}
\psfrag{y}{\small \hspace{-0.8cm} $(1-\hat{q}^2)^2 F(q^2)$}
\psfrag{f1}{\tiny $F_+$}
\psfrag{f2}{\tiny $F_0+\hat{q}^2F_T/(1+\hat{m}_\pi)$}
\includegraphics*[width=18pc, height=10pc]{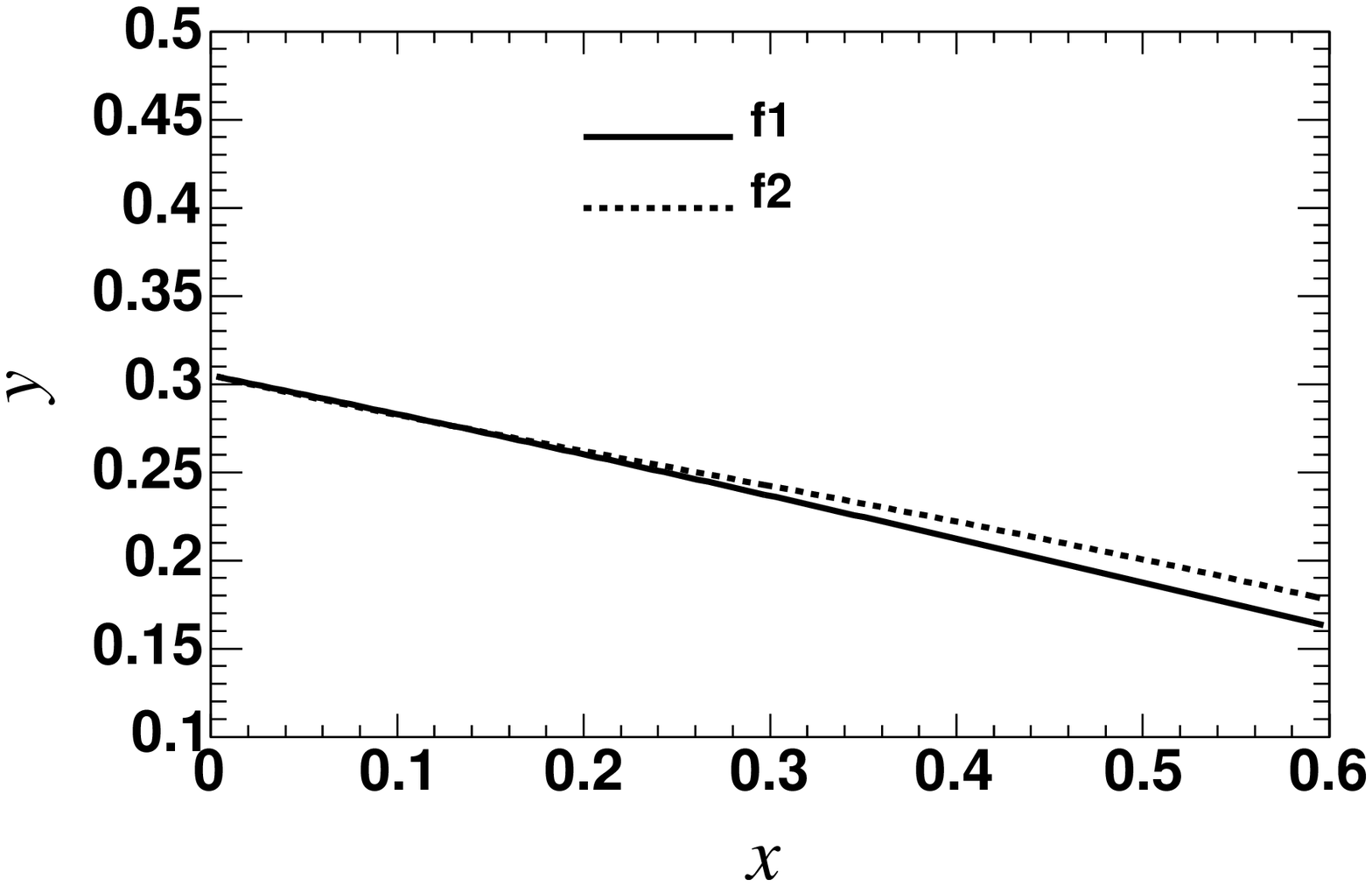}
\vspace{-16mm}
\caption{LCSR predictions for $B\to\pi$ form factors obeying
second-class symmetry relations.}
\vspace{-7mm}
\label{fig:P2}
\end{center}
\end{figure}
\begin{figure}[h]
\begin{center}
\psfrag{x}{\small $\hat{q}^2$}
\psfrag{y}{\small \hspace{-0.8cm} $(1-\hat{q}^2)^2 F(q^2)$}
\psfrag{f1}{\tiny $(1-\hat{q}^2) (V-A_2)/(1+\hat{m}_\rho)/2\hat{m}_\rho$}
\psfrag{f2}{\tiny $(1-\hat{q}^2)A_0$}
\psfrag{f2a}{\tiny \quad $ +\hat{q}^2(1-\hat{q}^2)(T_1-T_3)/2\hat{m_\rho}$}
\includegraphics*[width=18pc, height=10pc]{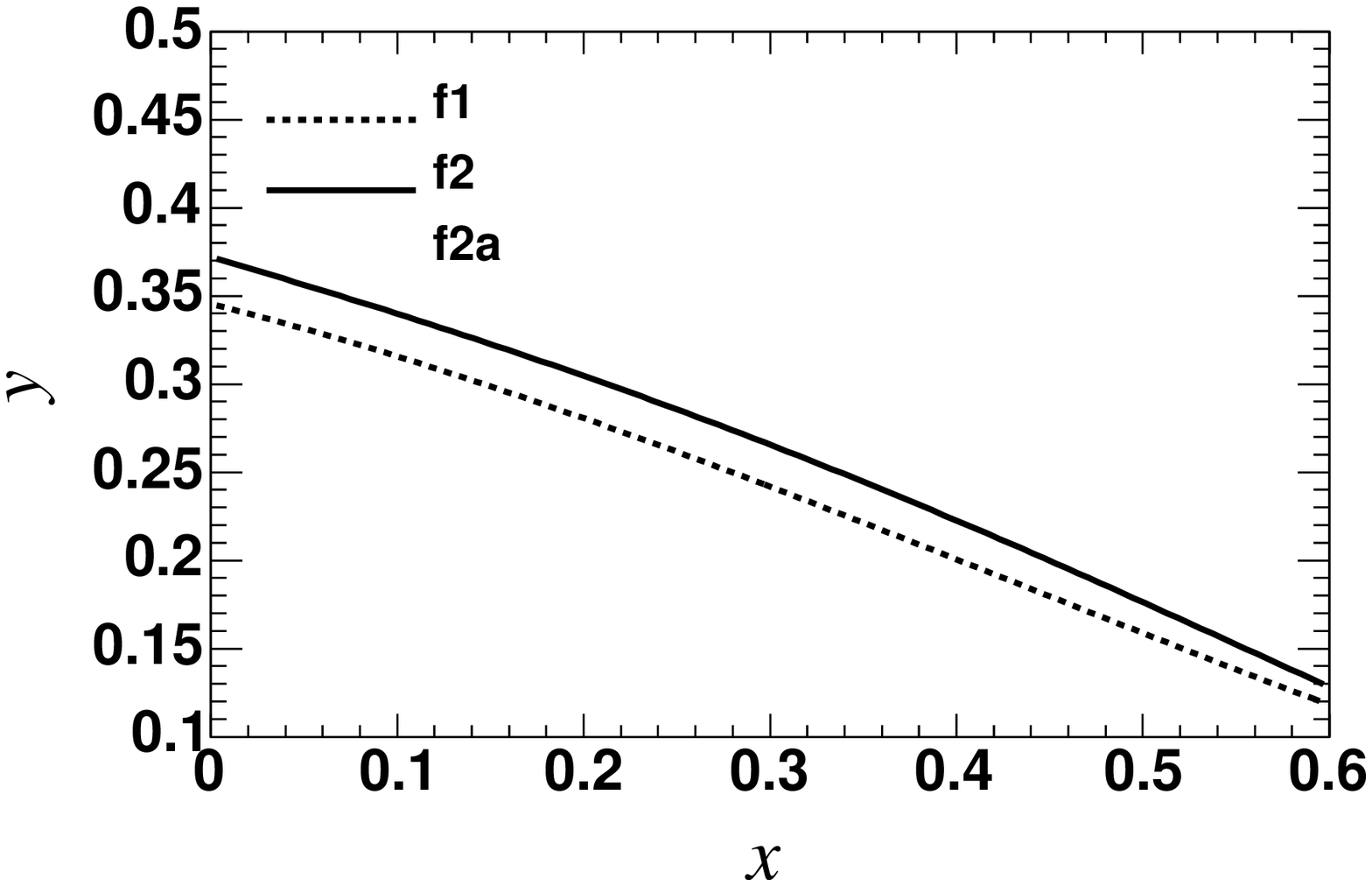}
\vspace{-16mm}
\caption{LCSR predictions for $B\to \rho_\parallel$ form factors obeying
second-class symmetry relations.}
\vspace{-8mm}
\label{fig:Vpar2}
\end{center}
\end{figure}

LCSRs give predictions for all $B\to \pi$ and $B\to \rho$ form factors at large recoil, 
and it is interesting to compare these predictions with the symmetry relations 
(\ref{eq:firstclass}), (\ref{eq:secondclass}) and (\ref{eq:thirdclass}).  
Figs.~\ref{fig:P}, \ref{fig:Vpar} and \ref{fig:Vperp} show the LCSR predictions for $B\to P$, 
$B\to V_\parallel$ and $B\to V_\perp$ decay form factors, respectively, using 
the results of \cite{Ball:1998kk,Ball:1998tj}.
Tensor form factors are evaluated at renormalization scale $\mu=m_b=4.8\,{\rm GeV}$.  
Normalizations are chosen such that at leading power, 
$F_i^{B\to M} = \zeta_M$ up to hard-scattering and radiative corrections.  
According to the power counting obtained in perturbation theory, the soft form factors scale as
$1/E^2\propto (1-\hat{q}^2)^{-2}$, and this factor is extracted in the plots. 

For the cases $B\to P$ and $B\to V_{\parallel}$, the form factors obey
only third-class relations, so that the deviation of the curves is a measure of 
hard-scattering terms.  In particular, since the hard-scattering corrections to 
$A_0$ and $V-A_2$ are positive, while that to $T_1-T_3$ is negative~\cite{Hill:2004if}, the deviations from
the symmetry relations may be particularly large for these curves. 
For the $B\to V_\perp$ case, both 
$A_1$ and $V$, and $T_1$ and $T_2$, obey first-class symmetry relations, whereas 
$A_1$ and $T_1$ obey only third-class relations.  
  
Figs.~\ref{fig:P2} and \ref{fig:Vpar2} show the form factors in (\ref{eq:secondclass}) 
obeying second-class symmetry relations.  
Comparison of the corresponding curves in Figs.~\ref{fig:P} and
\ref{fig:P2}, and in Figs.~\ref{fig:Vpar} and \ref{fig:Vpar2}, shows that in both cases the 
second-class relations are satisfied more accurately than the third-class 
relations, as expected.   

\section{Scaling violations}
\label{sec:renorm}

The LCSR results in Figs.~\ref{fig:P} - \ref{fig:Vpar2} show  
a significant deviation from the $1/E^2$ dependence 
predicted from naive power counting.  
In fact, using first and second class relations to isolate the 
soft-overlap terms, the sum rule results give in all cases, 
$d(\ln\zeta)/d\hat{q}^2 \approx 1.3$ at $q^2=0$, to be compared with the naive scaling 
prediction of $d(\ln\zeta)/d\hat{q}^2 = 2$.

Perturbative contributions to scaling violations can be investigated by considering the RG evolution 
equation~\cite{Bauer:2000yr,Becher:2003kh,Lange:2003pk},
\be
\label{eq:zetaRG}
  {d \ln \zeta(E,\mu)\over d\ln\mu} = - \Gamma_{\rm cusp}(\alpha_s) \ln{2E\over\mu} - \tilde{\gamma}(\alpha_s) \,, 
\ee
where at one-loop order, $\Gamma_{\rm cusp}(\alpha_s) = C_F\alpha_s/ \pi$ 
and $\tilde{\gamma}(\alpha_s)=-5C_F\alpha_s/ 4\pi$.  
The solution of (\ref{eq:zetaRG}) relates $\zeta(E,\mu)$ at different scales,
\be
\begin{aligned}
\label{eq:zetascales}
  &{\zeta(E,\mu)\over \zeta(E,\mu_0)}\! =\! 
  \left(2E\over \mu \right)^{a(\mu,\mu_0)} \\ 
  &\quad \times \exp\!\left[ S(\mu,\mu_0)\! -\! 
  \int_{\mu_0}^{\mu}\!{d\mu^\prime\over \mu^\prime}\tilde{\gamma}(\alpha_s(\mu^\prime)) \right] \,,
\end{aligned}
\ee
where
\be
\begin{aligned} 
  a(\mu,\mu_0) &= -\int_{\mu_0}^{\mu}\!{d\mu^\prime\over \mu^\prime}\Gamma_{\rm cusp}(\alpha_s(\mu^\prime)) \,, \\
  S(\mu,\mu_0) &= -\int_{\mu_0}^{\mu}\!{d\mu^\prime\over \mu^\prime}\Gamma_{\rm cusp}(\alpha_s(\mu^\prime))
  \ln{\mu\over\mu^\prime} \,.
\end{aligned}
\ee
For instance, up to hard-scale radiative corrections, $\zeta(E,\mu=2E)$ describes 
the soft-overlap part of the physical form factors,~%
\footnote{
In fact, taking the hard scale $\mu=2E$ (as opposed to $\mu=m_b$), the remaining dependence of the
Wilson coefficients on energy is very mild~\cite{Becher:2004kk}, so that e.g. $\zeta_P(E,\mu=2E)$ should accurately 
describe the energy-dependence of the soft-overlap contribution to $F_+$. 
}
and may be related to $\zeta(E,\mu_0)$, for a lower, energy-independent, scale $\mu_0$ (say $\mu_0= 1\,{\rm GeV}$).  
The slope at $q^2=0$ then satisfies: 
\be
\label{eq:softRG}
\begin{aligned}
  &{d\over d\hat{q}^2} \ln\zeta(E,\mu=2E) \bigg|_{q^2=0} 
  - {d\over d\hat{q}^2} \ln\zeta(E,\mu_0) \bigg|_{q^2=0} \\
  &\quad = \int_{\mu_0}^{m_b}{d\mu\over \mu} \Gamma_{\rm cusp}(\alpha_s(\mu)) + \tilde{\gamma}(\alpha_s(m_b)) \\
  &\quad \approx 0.28 - 0.13 \,,
\end{aligned}
\ee
where the last line is evaluated at $\mu_0=1\,{\rm GeV}$.  The first term on the right-hand side of (\ref{eq:softRG}) is 
positive for all $\mu_0<m_b$, while the second is independent of $\mu_0$.  Any large deviation of the form factor
slope from the naive scaling prediction, particularly any negative correction, must therefore arise from the 
nonperturbative function $\zeta(E,\mu_0)$.  

The situation is analogous to that for heavy-heavy meson transitions at large recoil. 
At leading order in the heavy-quark expansion, $B\to D$ form factors are described by the single
Isgur-Wise function, 
\be
  {\langle D_{v^\prime}| \bar{h}^{(c)}_{v^\prime} \Gamma h^{(b)}_{v} | B_v \rangle  \over \sqrt{ m_B m_D} }
  \!=\! -\xi(v\cdot v^\prime\!,\mu) {\rm tr}\!\left\{ \overline{\cal M}(v^\prime) \Gamma {\cal M}(v) \right\} \,,
\ee
which obeys the evolution equation, 
\be
\label{eq:xiRG}
  {d\ln \xi(v\cdot v^\prime,\mu)\over d\ln\mu}  = -\Gamma_{\rm cusp}( \phi, \alpha_s) \,. 
\ee
Here $\Gamma_{\rm cusp}(\phi,\alpha_s)$ is the universal cusp 
anomalous dimension~\cite{Korchemsky:wg}, 
and $\phi={\rm arccosh}(v\!\cdot\! v^\prime)$ is the angle between initial and final meson velocities. 
At one loop order,   
$\Gamma_{\rm cusp}( \phi, \alpha_s) = (C_F\alpha_s/\pi)\left( \phi\coth\phi - 1 \right)$.~% 
\footnote{
At large values of the cusp angle, the coefficient of $\ln v\!\cdot\! v^\prime$ is 
defined to be the angle-independent cusp anomalous dimension~\cite{Korchemskaya:1992je}, 
$\Gamma_{\rm cusp}(\alpha_s)$.  In the heavy-light case, the cusp angle becomes infinite, and the resulting RG equation 
takes the form (\ref{eq:zetaRG}), with a non-universal, but energy-independent, 
anomalous dimension $\tilde{\gamma}$.  
}
Considering large recoil $v\cdot v^\prime \sim m_b/\Lambda_{\rm QCD} \gg 1$,
the Isgur-Wise function $\xi$ behaves similarly to the soft SCET form factor $\zeta$.   
In particular, $\xi \propto (v\cdot v^\prime)^{-2}$ 
up to logarithms~\cite{Grozin:1996pq}.~%
\footnote{
Also in this case, competing ``hard-scattering'' terms enter at the same order.  At still larger 
recoil $ v\cdot v^\prime \gg  m_b/\Lambda_{\rm QCD}$, the hard-scattering terms dominate.
}
Perturbative scaling violations may again be calculated using (\ref{eq:xiRG}), but a nonperturbative 
dependence on $v\cdot v^\prime$ remains in $\xi(v\cdot v^\prime,\mu_0)$. 

\section{Summary}
\label{sec:conclusion}

Symmetry relations in the heavy-quark/large-energy limit for heavy-to-light meson form factors
provide a valuable handle on otherwise poorly understood hadronic parameters.  
Application of the symmetry relations 
to $D$-meson decays, especially into vector mesons, 
may be problematic due to large meson-mass effects; further experimental
investigation could clarify this.  Such effects may largely cancel in ratios relating $B\to M$ 
to $D\to M$ transitions at the same light-meson energy.  Here SCET generalizes corresponding HQET relations
to allow for large recoil energy, and would be important for relating the form factors near
maximum recoil for the semileptonic $D\to\pi$ decay.  

The analysis of semileptonic $B\to \rho$ decay becomes especially simple in the large-energy limit, where
only two helicity amplitudes contribute.  Extraction of the remaining form factors from processes such 
$B\to K^*\gamma$ or $B\to K^*l^+l^-$ (with an understanding of $SU(3)$ violations) or from $B\to\rho\gamma$ 
(with an understanding of the $q^2$ dependence of form factors) have potential to provide useful CKM constraints.  

A comparison of the symmetry relations to LCSR predictions shows no sign of symmetry-breaking power 
corrections beyond the $10\%$ level.  In addition, the LCSRs show a large correction to the perturbative 
scaling behavior of the soft form factors.  
A better understanding of such scaling violations from first principles is an important problem, as is
a study of power corrections for exclusive decays in the large-energy limit.   
Improved measurements coming from $B$ decays at BaBar and Belle, and $D$ decays at CLEO-c, will
provide numerous tests and applications of the SCET predictions.

\vspace{5mm}

\noindent{\bf Acknowledgments}
I thank T. Becher and M. Neubert for collaboration on projects underlying
the topics discussed here.   
Research supported by the Department of Energy under Grant DE-AC02-76SF00515.

\end{document}